# On the Proof of Vanishing Cosmological Constant in String Theory

Hirotaka Sugawara, OIST, Okinawa, Japan

The proof of the vanishing of the $N_0$-dimensional cosmological constant is given for the 10-dimensional supergravity–super–Yang–Mills theory. The supersymmetry is broken at will. In general, the 10-dimensional cosmological constant is not zero but the $N_0$-dimensional one is, where $N_0$ is the maximum dimension for which the vacuum values of fermion fields, $N_0$-components of the gauge fields and the antisymmetric field vanish. The proof is very simple and is mainly based on the "scale invariance" of the classical theory but it assumes a nouvelle interpretation of the quantum theory or, rather, that of the Planck constant.

We regard the Planck constant $\hbar$ not to be a constant but a field variable which is determined by the equation of motion. It is a super-invariant field thus it does not depend on the space-time variables. This interpretation is consistent with the fact that quantum gravity in string theory does not depend on the value of $\hbar$ due to the "scale invariance" of the classical theory. The proof utilizes the 10-dimensional super–Yang–Mills–supergravity formulation with possible stringy corrections. Purely stringy proof is left for the future work.

---

We start by writing down the 10-dimensional supergravity Langrangian with super–Yang–Mills fields [1].

$$L = \frac{-1}{2\kappa^2} \int e \left[ R + \frac{\phi}{2} \operatorname{tr}\left(F_{MN}^2\right) + 2\,(\partial_M \ln\phi)^2 + \frac{3\phi^2}{4} H_{MNP}^2 + \text{Fermion terms} \right] d^{10}x \,\text{---}\, (1).$$

Here, $e = \det\{g_{MN}\}$, $\phi$ is the dilaton field, $F_{MN}$ is the guage field and $H_{MNP}$

is the antisymmetric tensor field.
Fermions include the gravitino, dilatino and the gaugino fields.
We define the effective potential by the following equation:

$$e^{\frac{-i\hbar^{-1}}{2\kappa^2} \int e[R + 2V(\phi, F, H, g^{mn}, e\hbar^{-1})] d^{10}x} =$$

$$\int [d\phi dF dH \, d\chi d\lambda d\Psi d\bar{\chi} \, d\bar{\lambda} \, d\bar{\Psi}] \, e^{\frac{-i\hbar^{-1}}{2\kappa^2} \int e[R + 2V_0(\phi, F, H, g^{MN}, \chi, \lambda, \Psi)] d^{10}x} \text{ ---}$$

(2).

Here $\chi$ stands for the gaugino, $\lambda$ for the dilatino and $\Psi$ for the gravitino field respectively. The integration is done for the quantum fluctuations and the Boson fields on the left hand side is their vacuum expectation values. Note also that we do not integrate over the quantum fluctuations of the gravity field $g_{MN}$. Whether we should or should not include the quantum gravity contribution to the definition of the cosmological constant may be debatable. In this sense, the precise statement is that what we prove is the vanishing of the contribution of the matter fields with their quantum fluctuations to cosmological constant.

$$V_0 = \frac{\phi}{2} \, \text{tr}\left(F_{MN}^2\right) + 2 \, (\partial_M \ln\phi)^2 + \frac{3\phi^2}{4} H_{MNP}^2 + \text{Fermion terms} \text{ --- (3)}$$

A crucial observation is that (a) on the right hand side of equation (2), the dependence of $\hbar$ is in the form $e\hbar$ and that, (b) other than this factor of $e$, there is no explicit dependence on $g_{MN}$ except for the inplict dependence through the product such as $F_{MN}^a F^{aMN} = g^{KM} g^{LN} F_{KL}^a F_{MN}^a$. This shows that, if we choose the dependence to be $g^{MN}$ then there is no dependence on $g_{MN}$. Therefore, the classical $g^{MN}$ after the integration must be contracted with F or H. By assumption, the classical F and H have only m and n components where m, n indicate the compactified $10 - N_0$ components. This argument shows that $g_{\mu\nu}$ on the left hand side of equation (2) appears only through $e\hbar$ where $\mu$ and $\nu$ indicate the $N_0$ components.
(c) As is pointed out in reference [1] the right hand side of equation (2) is invariant under the following scale transformation:

$$g_{MN} \rightarrow t^{-2} g_{MN}, \quad \phi \rightarrow t^{-2} \phi, \quad \hbar \rightarrow t^{-8} \hbar,$$

$(\chi, \lambda, \Psi) \rightarrow \sqrt{t}\, (\chi, \lambda, \Psi)$ --- (4), if the integration measure $[d\phi dFdH\, dxd\lambda d\Psi d\bar{\chi}\, d\bar{\lambda}\, d\bar{\Psi}]$ is invariant under this transformation: For example we can put,

$$[d\phi dFdH\, dxd\lambda d\Psi d\bar{\chi}\, d\bar{\lambda}\, d\bar{\Psi}] = \prod_x \sqrt{\phi}\; d\phi dFdH\, dxd\lambda d\Psi d\bar{\chi}\, d\bar{\lambda}\, d\bar{\Psi}.$$

We now prove the following:
If the vacuuum values of the fermion fields $\chi$, $\lambda$, $\Psi$, $(A^a)_\mu$, $H_{\mu\nu\rho}$ and $g_{\mu m}$ with $\mu$, $\nu$ and $\rho$ less than or equal to $N_0 - 1$ and $m = N_0, \cdots, 9$, vanish, and if the non-zero vacuum value of $\phi$ does not depend on $x_M$, then we get $R_{N_0} = \lambda_{N_0} = 0$.

It is trivial to include the case of non-vanishing fermion pair condensations but we will not discuss it here to avoid complications. The essential ingradient is to assume that the effective potential on the left hand side of equation (2) is the potential not only for $\phi$, F, H but also for $\hbar$. Here $\hbar$ is assumed to be a supersymmetry singlet and, therefore, it has no space-time coordinate dependence from the outset. Note that one factor of $\hbar^{-1}$ is outside of the integral as usual, which may or may not imply that the Lagrangian to determine $\hbar$ is V rather than $\hbar^-$ V. First consider the case when the Lagrangian is V.
Thus we have two crutial equations of motion:

$$\frac{\partial}{\partial \phi} V\left(\phi,\, F,\, H,\, g^{mn},\, e\hbar^{-1}\right) = 0 \;---\; (5)$$

$$\frac{\partial}{\partial \hbar} V\left(\phi,\, F,\, H,\, g^{mn},\, e\hbar^{-1}\right) = 0 \;---\; (6)$$

Both are valid since both $\phi$ and $\hbar$ have no kinetic energy.
From equation (6) we get,

$$\frac{\partial}{\partial g_{\mu\nu}} V\left(\phi,\, F,\, H,\, g^{mn},\, e\hbar^{-1}\right) = \frac{\partial e}{\partial g_{\mu\nu}} \frac{\partial}{\partial e} V\left(\phi,\, F,\, H,\, g^{mn},\, e\hbar^{-1}\right) =$$

$$-\frac{\partial e}{\partial g_{\mu\nu}} \left(\frac{\hbar}{e}\right) \frac{\partial}{\partial \hbar} V\left(\phi,\, F,\, H,\, g^{mn},\, e\hbar^{-1}\right) = 0 \;---\; (7).$$

The scale invariance tells that
$t^2 V\left(\phi,\, F,\, H,\, g^{mn},\, e\hbar^{-1}\right) = V\left(t^{-2}\phi,\, F,\, H,\, t^2 g^{mn},\, t^{-2} e\hbar^{-1}\right)$
All the components are in the extra-dimensional space.

We differentiate this with respect to t and put t = 1. Then we get,

$$2V(\phi, F, H, g_{mn}, e\hbar^{-1}) = -2\phi \frac{\partial V}{\partial \phi} + 2g^{mn}\frac{\partial V}{\partial g^{mn}} + 2g^{\mu\nu}\frac{\partial V}{\partial g^{\mu\nu}} - 8\hbar \frac{\partial V}{\partial \hbar} \quad \text{---} \quad (8).$$

Substituting (5), (6) and (7) to (8), we get,

$$V(\phi, F, H, g_{mn}, e\hbar^{-1}) = g^{mn}\frac{\partial V}{\partial g^{mn}} \quad \text{---} \quad (9).$$

We now write down the gravity equation:

$$R_{KL} - \frac{1}{2}\left[R + V(\phi, F, H, g_{mn}, e\hbar^{-1})\right] g_{KL} + g_{KM} g_{LN} \frac{\partial V(\phi, F, H, g_{mn}, e\hbar^{-1})}{\partial g_{MN}} = 0 \quad \text{---} \quad (10).$$

For $\mu, \nu = 0, 1, \cdots, N_0 - 1$, this becomes,

$$R_{\mu\nu} - \frac{1}{2}[R + V] g_{\mu\nu} = 0 \quad \text{----} \quad (11).$$

Here we used equation (7) and $g_{\mu m} = 0$ by assumption.

Since we have $R = R_N + R_{10-N}$ for the case $g_{\mu m} = 0$, we have

$$R_{\mu\nu} - \frac{1}{2}[R_{N_0} + R_{10-N_0} + V] g_{\mu\nu} = 0$$

This shows that the effective $N_0$-dimensional cosmological constant is given by:

$$\lambda_{N_0} = R_{10-N_0} + V \quad \text{---} \quad (12)$$

The trace of this equation is,

$$R_{N_0} - \frac{N_0}{2}[R_{N_0} + R_{10-N_0} + V] = 0,$$

or using equation (12),

$$\left(1 - \frac{N_0}{2}\right) R_{N_0} - \frac{N_0}{2} \lambda_{N_0} = 0 \quad \text{---} \quad (13).$$

The extra-dimensional components of equation (10) reads,

$$R_{kl} - \frac{1}{2}[R + V] g_{kl} + g_{km} g_{ln} \frac{\partial V}{\partial g_{mn}} = 0,$$

and the trace of this equation is,

$$R_{10-N_0} - \frac{10 - N_0}{2}[R_{N_0} + R_{10-N_0} + V] + g_{mn}\frac{\partial V}{\partial g_{mn}} = 0 \quad \text{---} \quad (14).$$

Substituting equation (9) to equation (14) and using equation (12), we get,

$$\lambda_{N_0} - \frac{10 - N_0}{2} [R_{N_0} + \lambda_{N_0}] = 0 \text{ --- (15)}.$$

From equation (13) and equation (15) we finally get,

$$R_{N_0} = \lambda_{N_0} = 0 \text{ --- (16)},$$

Next we assume the Lagrangian is $\hbar^{-1} V$.

In this case, we still have equation (5) but equations (6) and (7) are replaced by,

$$-\hbar \frac{\partial}{\partial \hbar} V(\phi, F, H, g^{mn}, e\hbar^{-1}) + V = 0 \text{ --- (6)}', \text{ and,}$$

$$\frac{\partial}{\partial g^{\mu\nu}} V(\phi, F, H, g^{mn}, e\hbar^{-1}) = \frac{1}{2} g_{\mu\nu} V \text{ --- (7)}'$$

The gravity equations also get modified as,

$$R_{\mu\nu} - \frac{1}{2} [R_{N_0} + R_{10-N_0}] g_{\mu\nu} = 0 \text{ ---- (11)}'$$

and

$$R_{10-N_0} - \frac{10 - N_0}{2} [R_{N_0} + R_{10-N_0}] = 0 \text{ --- (14)}'.$$

To obtain (14)' we have used the scaling equation (8) together with equations (5), (6)' and (7)'.

From equations (11)' and (14)' we derive,
$R_{N_0} = \lambda_{N_0} = R_{10-N_0} = 0$.

Q.E.D.

Therefore, the either choice of Lagrangian leads to the vanishing cosmological constant.

Several comments are in order:
(1) One can check explicitly that, in case $V = V_0$ (no fermion term) =

$$\frac{\phi}{2} \text{tr}(F_{MN}^2) + 2(\partial_M \ln\phi)^2 + \frac{3\phi^2}{4} H_{MNP}^2 \text{ --- (17)},$$

what we have proved is correct i.e. $R_{N_0} = \lambda_{N_0} = 0$.
But, in this case, equation (5) gives,

$$\phi = \frac{-\text{tr}(F_{MN}^2)}{3 H_{MNP}^2} < 0 \text{ --- (18)}.$$

This shows that the gauge field is a ghost. Therefore, higher order correction is needed to make this value positive.

(2) The string correction in the tree level does not change our result as long as the same assumption

of vanishing fermion and the vanishing $N_0$-component vacuum values are made because the ptential is $\hbar$-independent. If the scale invariance remains correct, we can include all the quantum fluctuation of the matter fields. But the stringy proof may be appropriate in this case.

(3) The most mysterious is equation (6) and the claim is that it determines the value of $\hbar$. One may legitimately ask the question, "In terms of what?". The answer is, "In terms of the size of extra-dimensional space".

Assume $N_0 = 4$, then there is no equation to determine the value of $V_6$, where $V_6$ denotes the volume of the 6-dimensional compactified space. Suppose that the size $V_6^{\frac{1}{6}}$ is $n_0 \, l_p$ where $l_p$ is the usual Planck length, then,

$$V_6^{\frac{1}{6}} = n_0 \, l_p = n_0 \sqrt{\frac{\hbar G}{c^3}} = n_0 \sqrt{\frac{\hbar \kappa^2}{8 \pi V_6 \, c^3}} \quad --- (19),$$

where we have used,

$$\kappa_4 = \sqrt{8 \pi G} = \sqrt{\frac{1}{V_6}} \, \kappa \quad --- (20).$$

Solving (19) for $\hbar$, we get,

$$\hbar = \frac{8 \pi \, (V_6)^{\frac{4}{3}} \, c^3}{n_0^2 \, \kappa^2} \quad --- (21).$$

In principle, $n_0$ is determined by equation (6). Of course this not mean $V_0$ is a fundamental constant replacing $\hbar$. It is a boundary value of the theory which does not appear in the action.

(4) Why $N_0$ is equal to 4 remains a mystery. The analysis of the equations to determine the vacuum values of F and H $\left(\text{in lowest order they are } \partial_p [e H^{pmn}] = 0,\right.$ and $\left.\partial_P [e F^{cPQ}] - e f^{abc} F^{aPQ} A_P^b + \frac{3}{2} e \phi H^{QRS} F^c_{RS} = 0\right)$ must provide the clue to this problem.

(5) The problem of "dark energy" of our universe should not be confused with the issue of vanishing cosmological constant. Our universe is not a vacuum. It is now believed to have started with the "inflationary stage" before the "big bang". Dark energy should be understood as a remnant of the inflationary energy which did not go into the ordinary matter energy. Such a model was presented by the present author in reference [2]. In this model, two instanton potentials, one for E(6) and one for SU(3), exist in the universe which started at certain point on either E(6) side or the SU(3) side and, at the present stage of the universe, it is close to the bottom of the sum of two potentials. It is no mystery in this kind of model that the amount of dark energy is not far from that of ordinary matter energy since the orgin of the energy is the same.

(6) To convince ourselves that there is no problem of cosmological constant in the so-called " standard model" or in its supersymmetric extension, we have to be able to prove that it can be derived from a Lagrangian such as written in equation (1). The work toward this goal is under way.

(7) The above argument implies that other so-called fundamental constants like the velocity of light c or the Newton's constant should be treated as dynamical variables just as $\hbar$ and the string theory seems to be the right framework to do it. Light velocity is already the vacuum value of $g_{00}$ which is, of course, a dynamical variable. This issue will be left to the future work.

(Acknowlegement) This work was started when the author was visiting UC Santa Cruz in 2012 and was almost finished when he was visiting UCLA in 2013. He is indepted to Professor Michael Dine of UC Santa Cruz and Professor Roberto Peccei of UCLA for their hospitality. He is also thankful to Professors Jiro Arafune and Yoshimitsu Shimizu for discussions. He acknowledges Jonathan Dorfan, President of OIST for my leaves of absence from my administrative duties at OIST during my visits to UCSC and UCLA.